\documentclass[bibyear]{aa}
\pdfoutput=1
\usepackage{amsmath, amsfonts, amssymb}
\usepackage{txfonts}
\usepackage{graphicx}
\usepackage{lipsum}
\usepackage{natbib}
\bibpunct{(}{)}{;}{a}{}{,} 
\authorrunning{Menci et al.}
\titlerunning{Early growth of massive black holes and dynamical dark  energy}

\begin{document}

\title{Early growth of massive black holes in dynamical dark energy models with negative cosmological constant}

\author{N. Menci\inst{1}
\and M. Castellano\inst{1}
\and P. Mukherjee\inst{2}
\and D. Roberts\inst{3}
\and P. Santini\inst{1}
\and A.A. Sen\inst{2}
\and F. Shankar\inst{3}
}

\institute{Istituto Nazionale di Astrofisica (INAF), Osservatorio Astronomico di Roma, Via Frascati 33, 00078 Monte Porzio Catone (RM), Italy
\and 
Centre for Theoretical Physics, Jamia Millia Islamia, New Delhi-110025, India
\and School of Physics \& Astronomy, University of Southampton, Highfield, Southampton SO17 1BJ, UK
}

\abstract{Recent results from combined cosmological probes indicate that the Dark Energy component of the Universe could be dynamical. The simplest explanation envisages the presence of a quintessence field rolling into a potential, where the Dark Energy energy density parameter $\Omega_{DE}=\Omega_{\Lambda}+\Omega_{x}$ results from the contribution of the ground state energy $\Omega_{\Lambda}$
 and the scalar field energy $\Omega_{x}$. Provided that $\Omega_{DE}\approx 0.7$, negative values of $\Omega_{\Lambda}$ can be consistent with current measurements from cosmological probes, and could help in explaining the large abundance of bright galaxies observed by JWST at $z> 10$, largely exceeding the pre-JWST expectations in a $\Lambda CDM$ Universe. }{Here we explore to what extent such a scenario can account also for the early presence of massive Black Holes (BHs) with masses $M_{BH}\gtrsim 10^7\,M_{\odot}$ observed at $z\gtrsim 8$, and for the large over-abundance of AGN with respect to pre-JWST expectations. 
 Our aim is not to provide a detailed description of BH growth, but rather to compute the maximal BH growth that can occur in cosmological models with negative $\Omega_{\Lambda}$ under the simple assumption of Eddington-limited accretion onto initial light Black Hole seeds with mass $M_{seed}\sim 10^2\,M_{\odot}$ originated from PopIII stars. }{To this aim we develop a simple analytic framework to connect the growth of dark matter halos to the maximal growth of BHs within the above assumptions.}{We show such models can account for present observations assuming values of $\Omega_{\Lambda}\approx -1$, simultaneously boosting both galaxy and AGN number counts without invoking any additional physics. This would allow us to trace the observed excess of bright and massive galaxies and the early formation of massive Black Holes and the abundance of  AGN to the same cosmological origin.}{}
\maketitle

\section{Introduction}
\label{sc:introduction}

In recent years, major observational breakthroughs, largely provided by the James Webb Space Telescope (JWST), have severely challenged the canonical $\Lambda$CDM scenario for the formation of cosmic structures. 
These observations revealed a sensible excess of massive galaxies at $z\gtrsim 6$ \citep{Labb__2023,xiao2024accelerated,Casey_2024ApJ} and of star forming, UV bright galaxies at $z\gtrsim 10$ \citep{Castellano_2022,Finkelstein_2022,Atek_2022,10.1093/mnras/stac3535,Robertson_2024,
Finkelstein_2023,perez_2023,Donnan_2022,Harikane_2023,Bouwens_2023a,McLeod_2023a,
adams24,Finkelstein_2024,perez2025} over the pre-JWST models of galaxy formation in the $\Lambda$CDM scenario.

Such a fast growth of the galaxy stellar mass content of galaxies seems to be accompanied by an even faster growth of the mass $M_{BH}$ of the Supermassive Black Holes (BHs) hosted in high redshift galaxies, which is thought  to originate from the accretion of gas onto initial BH seeds, see \cite{Volonteri2021Nat} for a review). BH masses as large as $M_{BH}\approx 10^{7-8.5}\,M_{\odot}$, as inferred from the broadening of Balmer lines, have now been found already at $z\sim 6-8.5$ \citep{Kokorev_2023,Harikane2023ApJ,Maiolino2024a,Furtak2024Nature,Juod_balis_2024},
while at higher redshifts $z\gtrsim 10$ the observations 
of high-ionization lines \citep{Maiolino2024a} and X-ray detections \citep{Bogdan_2024NatAs,Kovacs_2024ApJ} have led to infer BH masses as high as $M_{BH}=10^{6.2-7.9}\,M_{\odot}$. 

These high masses pose a crucial challenge for theoretical models of BH seeding and growth due to the lack of cosmic time available for their assembly, see reviews by \cite{inayoshi2024,Fan_2023ARAA}. For instance, the assumption that the initial BH seeds are constituted by the remnants of metal-free PopIII  with  mass $M_{seed}\sim 10^2\,M_{\odot}$ \citep{Carr1984MNRAS,Abel_2002Sci} and that the accretion is Eddington limited, seems to be excluded since in the $\Lambda$CDM framework there is not sufficient time to build up such high measured BH masses by $z\approx 8$, even assuming continuous accretion throughout the available cosmic time. 

In addition to the large masses of high-redshift BHs, the demography of the BHs in  accreting phase (the Active Galactic Nuclei, AGN) is also at variance with pre-JWST expectations \citep{HaricaneAGN2023ApJ,Maiolino2024A&A,Juod2025,Scholtz2025A&A,Akins2024}, with a  number density of AGNs exceeding the pre-JWST measurements of the quasar abundance by more than one order of magnitude, at least at intermediate luminosities. Indeed, the AGN population revealed by JWST includes previously unseen, heavily reddened and extremely compact ($R_e\lesssim 100 $ pc) objects at $4\lesssim z\lesssim 10$, known as “little red dots” \citep{Matthee2024,Furtak2023,Labbe2025ApJ,Kokorev_2024,Greene2024ApJ,Akins2024,Kocevski2025ApJ}. Their compact size and peculiar spectra suggest that their emission is presumably dominated by obscured active galactic nuclei (AGN), an interpretation supported by the detection of broad emission lines in candidates possibly spectroscopically confirmed (and even by a direct measurement of the accreting BH mass, see \cite{Juod2025}), although alternative explanations exist  \cite{Ananna_2024} which suggest that their properties might be - at least partially - explained in terms of stellar emission \cite{Baggen_2024}.

Within the context of $\Lambda$CDM, all of the above observations  call for the presence of different additional astrophysical processes that must be at work at high redshift $z\gtrsim 8$. While several approaches have been proposed to produce an accelerated early phase of star formation and UV emission with respect to the pre-JWST expectations - like stochastic star formation \citep{shen2023impactuv,Sun_2023,Gelli_2024}, reduced dust attenuation \citep{Ferrara2024a}, lower metallicity stellar populations or a top-heavy IMF \citep{trinca_stae651,Yung2023}, clumpy star formation \citep{somerville2025MNRAS} and/or weaker supernova feedback \citep{Dekel_2023} - 
the presence of massive BH at high redshift $z\gtrsim 8$ and the large number density of AGN have been generally explained in terms of additional assumptions for the initial BH seeds and for the accretion properties of AGN. Theoretical models that have been employed to investigate the high-redshift growth of BHs \citep{TrincaAGN,Porras2025,LaChance2025,Camelli2024,DayalMaiolino,mcclymont2025,Quadri2025} can simultaneously reproduce the over-massive nature and high space densities of the observed AGN by invoking episodic super-Eddington accretion \citep{Madau2014ApJ,Volonteri_2015}, and/or assuming the existence of heavy seeds with masses $M_{seed}\sim 10^5\,M_{\odot}$ \citep{Bromm_2003,Begelman_2006}. Such seeds should form from the direct collapse of gas clouds in atomic-cooling halos with virial temperatures $T\sim 10^4$ K characterized by a pristine gas composition to avoid fragmentation and irradiated by a Lyman-Werner  radiation intense enough to dissociate $H_2$ molecules so as to avoid star formation. Their formation thus  requires rather peculiar conditions which are relatively rare, see, e.g., \cite{Shauer2017,Latif_2022}. As for the super-Eddington accretion, its  plausibility and effectiveness within realistic conditions provided by the large-scale cosmological environment is subject to intense ongoing studies, see, e.g., \cite{Zhu2022,Ni2022MNRAS,Bhowmick2022,Lupi2024}).

All the above conditions for  star formation and  BH growth at high redshift require substantial modification of the pre-JWST picture of galaxy formation of SMBH evolution at high redshift $z\sim 10$, while essentially preserving at lower redshifts the canonical picture of gradual gas conversion into stars and of BH growth through Eddington limited accretion, see, e.g.,\cite{lai2024smbh}. 

However, an alternative possibility is that the tension between the abundance and masses of high-redshift galaxies and BHs observed by JWST and the $\Lambda$CDM predictions originates in the assumed cosmological model.  
While the need for the introduction of substantial change or for more  exotic physics affecting the early evolution of galaxies and BHs does not constitute on its own a conclusive evidence that the underlying cosmological model has to be revised, a growing body of independent observational evidence from cosmological probes suggests that the standard cosmological model, with dark energy (DE) described by a cosmological constant $\Lambda$, may be incomplete. In particular, the recent measurements \citep{Adame_2025,LodhaPhysRev,Calderon_2024,DESI:2025zgx,andrade2025} of  baryon acoustic oscillations (BAO)  by the Dark Energy Spectroscopic Instrument (DESI), when complemented with Planck cosmic microwave background (CMB) data and type Ia supernovae (SNIa) distance moduli measurements \citep{Scolnic_2022,Brout_2022,rubin2025,Abbott_2024,Sanchez_2024,Vincenzi_2024} yield evidence for a time-evolving, dynamical dark energy (DDE) \citep{Adame_2025,DESI:2025zgx,lodha2025}, with a confidence level ranging from $\sim 3\sigma$ to $4\sigma$ depending on the SNIa sample used in the constraints \citep{Adame_2025}. The preference for DDE models compared to $\Lambda$CDM is robust with respect to the different parameterizations for the time evolution of DE \citep{lodha2025,Giare_2024}.

The most straightforward interpretation of these results 
is that the DE is provided by the dynamics of a quantum field $\phi$  rolling over  a potential $V(\phi)$, which would yield a time-changing  equation-of-state parameter $w=[\dot\phi^2/2-V(\phi)]/[\dot\phi^2/2+V(\phi)]$ \citep{Peebles:1987,Ratra:1987rm,Wetterich:1994bg,Caldwell:1997ii}. The vacuum state of such a field 
corresponds to a cosmological constant $\Lambda$, while the total DE density $\rho_{DE}=\rho_x+\rho_{\Lambda}$ is contributed by a dynamical part with energy density $\rho_x$ and by the contribution of the ground state $\rho_{\Lambda}$. The observed low-redshift ($z\lesssim 1$) accelerated expansion of the Universe \citep{SupernovaSearchTeam:1998fmf,SupernovaCosmologyProject:1998vns} only requires  the total DE density $\rho_{DE}$ to be positive, leaving open the possibilities for positive and negative values of $\Lambda$. Although the commonly explored combinations of cosmic parameters generally assume positive $\rho_{\Lambda}$, corresponding to a de Sitter (dS) vacuum, 
the resulting scenarios have a number of drawbacks: e.g., constructing a stable, positive vacuum energy (a de Sitter vacuum) is notoriously difficult within string theory \citep{Danielsson:2018ztv,Vafa:2005ui,Palti:2019pca,Grana:2021zvf}; in addition, when $\Omega_{\Lambda}>0$ is assumed, the present cosmological constraints yield phantom-crossing behavior of the equation-of-state parameter $w$ which evolves from values $w\leq -1$ at at $z\gtrsim 1$ to values $w\geq -1$ at later times, thus implying a phantom phase which would imply a negative kinetic contribution $\phi^2$ to the equation of state parameter $w$ \citep{Ludwick}, and violate the null energy condition 
\citep{Vikman:2004dc,Carroll:2004hc,Nojiri:2005sx,Oikonomou:2022wuk,Trivedi:2023zlf}.

On the other hand,
cosmological models characterized by a composite DE sector with a negative cosmological constant (NCC) $\Lambda<0$
 -  corresponding to potentials $V(\phi)$ featuring a negative minimum, i.e., an anti-dS (AdS) vacuum -  appear naturally within string theory, see \cite{Maldacena:1997re}; specific string-inspired models with a NCC are described in \cite{Murai2025,svrcek2006,luu2025DDE}. In fact, 
they have been proposed as attractive alternatives that can satisfy the late-time acceleration constraints still remaining consistent with the other standard cosmological observations 
\citep{Cardenas:2002np,Poulin:2018zxs,Dutta:2018vmq,Visinelli:2019qqu,Ruchika:2020avj,DiValentino:2020naf,Calderon:2020hoc,Sen:2021wld,Malekjani:2023ple,Adil:2023exv,mukherjee2025quintessentialimplicationspresenceads,WangPHR}). 
Indeed,  recent studies by some of us have  shown that such models can enhance high-redshift structure formation, making them promising candidates for explaining the JWST galaxy excess \citep{Adil:2023ara,Menci_2024,menci2024excess,chakraborty2025}. 

Motivated by such findings, here we explore the impact of assuming a DDE cosmology with a negative cosmological constant (NCC) on the early formation of massive BHs and on the demography of AGN. Specifically, we investigate whether NCC models can  alone account  for the recent JWST observations without introducing super-Eddington accretion phases or an initial population of heavy BH seeds, to assess whether such cosmological scenarios could constitute a viable, unified solution to all the issues raised by the high-$z$ observations by JWST. Following the same line in our earlier works, we adopt a simple description to relate the stellar properties of galaxies to their DM halo, which is known to provide an excellent statistical description of the galaxy population at 
$z< 8$, while assuming initial BH seeds with mass $M_{seed}\approx 10^2\,M_{\odot}$  originating from PopIII stars, and Eddington limited accretion. Under such conditions, we then compute the  maximal growth (corresponding to continuous accretion at the Eddington rate) of massive BHs assuming NCC cosmologies, and compute the corresponding maximal masses of high-redshift SMBHs and the impact on the AGN demography. The paper is organized as follows: In Sect. 2 we describe the NCC DE models and their impact on the Universe expansion and on the growth of cosmic structures. In Sect. 3 we recall how the abundance of DM halos is derived in such cosmological models, and how we relate the DM halo mass of the host galaxies to the SMBH masses and to the AGN accretion rate assuming continuous accretion at the Eddington limit and BH seeds with mass $10^2\,M_{\odot}$. Based on such a model, in Sect. 4 we compare the growth of BH masses and the AGN abundance at $z\gtrsim 5$ with recent JWST observations. Sect. 5 is devoted to discussion, while in Sect. 6 we present our conclusions. 

\section{Dark energy models}
\label{sec:ncc}

We consider a DE sector consisting of a cosmological constant $\Lambda \gtrless 0$  with associated vacum energy density $\rho_{\Lambda}=\Lambda/8\pi G$ which can be positive or negative, corresponding to a se Sitter (dS) or Anti-de Sitter (AdS) vacuum respectively. On top of this we consider a DE component with energy density $\rho_x(a)>0$ which evolves with the expansion factor $a=1/(1+z)$. Rather than focusing to a specific  model for the evolving DE component, we parametrize the time evolution of its equation-of-state (EoS) parameter $w_x(a)$ using the widely adopted Chevallier-Polarski-Linder (CPL) form~\citep{Chevallier:2000qy,Linder:2002et}:
\begin{eqnarray}
w_x(z) = w_0 + w_a(1-a).
\label{eq:wxzcpl}
\end{eqnarray}
where $w_0$ describes  the  behavior  of DE in the local Universe, the value of $w_a$ describes its evolution back in cosmic time.

This parametrization captures the EoS behavior in several physical DE models as discussed  in \cite{Linder:2002et,Linder:2006sv,Linder:2007wa,Linder:2008pp,Scherrer:2015tra}) (see also\cite{Perkovic:2020mph} for a comparison among different parametrization proposed in the literature), and allows us to compare with earlier works.

The  energy density of the evolving DE component then evolves according to the following:
\begin{eqnarray}
\rho_x(a)=\Omega_x\rho_{\rm c0}a^{-3(1+w_0+w_a)}e^{ \left [ -3w_a\,(1-a) \right ] }\,. \nonumber \\
\label{eq:rhoxz}
\end{eqnarray}
where $\Omega_x \equiv \rho_x(1)/\rho_{\rm c0}$ is the density parameter of the evolving DE component and  $\rho_{\rm c0}$ is the critical density. 
Assuming a spatially flat Friedmann-Lema\^{i}tre-Robertson-Walker Universe, the evolution of the Hubble rate (with respect to its present value $H_0$) in the matter-dominated era is governed by the following equation:
\begin{eqnarray}
E^2(a)\equiv H^2(a)/H_0^2=\Omega_m\,a^{-3} + \Omega_{\Lambda} + \Omega_xa^{-3(1+w_0+w_a)} e^{ \left [ -3w_a (1-a) \right ] }
\label{eq:fried}
\end{eqnarray}

where $\Omega_m$ is the matter density parameter, $\Omega_{\Lambda} \equiv \Lambda/3$ is the  density parameter associated with the cosmological constant, and $\Omega_m+\Omega_{\Lambda}+\Omega_x=1$ holds. The density parameter of the combined DE sector is $\Omega_{\text{DE}} \equiv \Omega_x+\Omega_{\Lambda}$. Although $\Omega_{\Lambda}$ can be negative, the density parameter associated to the \textit{total} DE density has to be $\Omega_{\text{DE}} \approx 0.7$
in order to be able to drive the observed cosmic acceleration at low redshift $z\lesssim 1$. 

We notice that, in order to agree with observations, more negative values of $\Omega_{\Lambda}$ need to be compensated by more negative values of $w_x$, moving towards the phantom regime ~\citep{Visinelli:2019qqu,Sen:2021wld,Adil:2023ara}. These considerations lead to upper limits of order unity in $\vert \Omega_{\Lambda} \vert$, corresponding to a negative vacuum energy density  $\lesssim 10^{-123}$ in Planck units. Although this may seem to represent the usual problem related to the small absolute value of the cosmological constant, recent studies 
~\citep{Demirtas:2021nlu,Demirtas:2021ote} have shown that supersymmetric AdS$_4$ vacua of the above magnitude naturally arise in the framework of string theory. 

\section{Method}
\label{sec:ncc}

We build a flexible and transparent analytic model that can provide upper limits to the growth of galaxies and their central supermassive BHs associated to DM structure formation in a dynamical dark energy framework. The steps we follow in our procedure are the following:
\begin{itemize}
    \item We first generate a sample of host halo masses extracted from the halo mass function at an initial redshift $z_i=20$.
    \item We then grow each central DM halo above 3.5$\sigma$ of the primordial density field along its main progenitor branch using analytical recipes 
    \item Within each DM halo we grow a BH via mergers and accretion. An upper limit on the growth of BHs due to mergers is obtained assuming that the fractional increase via mergers is directly linked to the fractional growth in host DM mass, while the growth via accretion is assumed to be a continuous process at the Eddington limit.
    \item From our mock sample of central DM haloes and their BHs we build the mean BH mass-halo mass relation which we then use to convert the halo mass function into a BH mass function. 
    \item From the BH mass function we then derive the AGN luminosity function, a straightforward step as in our approach all BHs are continuously accreting (duty cycle equal to unity) at the Eddington limit. 
    \item For completeness, in what follows we will also compute from our models the implied galaxy UV luminosity function at $z\gtrsim10$. 
\end{itemize}
The last step allows us to make a direct comparison with current data on the $z\sim 5-6$ AGN luminosity function, although the most revealing comparisons will be the ones at $z\gtrsim7$. We note that in principle it would be more straightforward to carry out a direct comparison with the BH mass function, but current estimates are mostly available at $z<6$, a time when the effects of a dynamical dark energy wear off. We also stress that for the conversion from halo to BH mass function we do not include any scatter as we are already assuming a strictly \textit{maximal} growth for each BH, given that in our modelling even the tiniest growth in the DM host halo induces a parallel growth in the central BH, and the latter further grows via constant Eddington-limited accretion. We remind that we are not including any super-Eddington accretion as our aim is to understand whether a strictly Eddington-limited accretion starting from PopIII BH seeds could be a viable model in matching the newest high-$z$ AGN data in the context of NCC cosmologies.  

\subsection{Abundance of collapsed objects}
\label{sec:ncc}

The abundance of collapsed objects in the Universe is described by the  mass function $dN(M)/dM$. We briefly recall its computation in NCC cosmologies,  computed for different values of $\Omega_{\Lambda}$ following the lines in \cite{Menci2022}. 

We base on the Press \& Schechter approach 
 \citep{Peacock1999_Textbook, Padmanabhan2002_TextbookVol3, Dodelson2020_Textbook} that relates the mass distribution to the properties of the linear perturbation density field. These are enclosed in the  
  the variance $\sigma(M,a)$ of matter density fluctuations smoothed on the comoving scale corresponding to the mass scale $M$.  In terms of the linear power spectrum of matter density fluctuations $P_{L}(k,a)$ computed at cosmic expansion factor $a$, the 
  variance writes 
\begin{eqnarray}
\sigma^2(M,a)=\frac{1}{2\pi^2}\int_0^\infty dk\,k^2 P_L(k,a) {W}^2(k, R)
\label{eq:sigma}
\end{eqnarray}
where ${W}(k, R) = 3 [\sin(kR) - kR \cos (kR)] /(kR)^3$ is the Fourier transform of the real-space spherical top-hat window function of radius $R
 = \left(3M/4\pi \bar{\rho}\right)^{1/3}$, and $\bar{\rho}$ is the mean comoving background matter density. The linear power spectrum $P_L(k,z)$ of matter density fluctuations as a function of wavenumber $k$ at a given expansion factor $a$ can further be expressed as:
\begin{eqnarray}
P_L(k,z)=P_0 k^{n_s} T^2(k) D^2(a)
\label{eq:pk}
\end{eqnarray}
where $P_0$ is a normalization constant, which is fixed using the present-day mass variance ($\sigma_8$) on a scale of 8 $h^{-1} Mpc$, $T(k)$ is the CDM matter transfer function, and $D(a)$ is the linear growth factor 
 normalized to 1 at present cosmic time.
Assuming that halos form when the linear perturbations reach critical linear overdensity $\delta_c$ for collapse, the mass function takes the  general form 

\begin{equation}
{dN \over dM}={\overline{\rho}\over M^{2}}\,{d ln\, \nu\over d\,\,ln M}\,
\nu\,f(\nu)
\label{eq:dndm}
\end{equation}

where $\nu=\delta_c/\sigma(M,a)$ is the critical height of density perturbations relative to the r.m.s. value of the perturbations. 
In the following we adopt the Sheth and Tormen  \citep{ST99}  mass function 
\begin{equation}
\nu\,f(\nu)=A\Bigg({1 \over \overline{\nu}^{2q}}+1\Bigg)\,  {\overline{\nu}^{2}\over \pi}     e^{-\overline{\nu}^{2}/2} 
\end{equation}
where the parameters  $A=0.32$, $a=0.71$ and $q=0.3$  are related to the physics of collapse. 

The growth factor $D(a)$ plays a major role in determining the extension to large masses of the mass distribution. The faster is the growth of density perturbations, the larger is the probability that rare, high-mass overdensities  reach the threshold for collapse. The evolution of the growth factor is governed by the following equation (see, e.g., \cite{Adil:2023ara} and references therein).

\begin{eqnarray}
\delta^{\prime\prime} + \left [ \frac{3}{a}+\frac{E^\prime(a)}{E(a)} \right ]\delta^{\prime} -\frac{3}{2}\frac{\Omega_m}{a^5 E^2(a)}\delta=0\,,
\label{eq:linear_growth}
\end{eqnarray}
where $^\prime$ indicates a derivative with respect to the scale factor $a$, and $E(a) \equiv H(a)/H_0$ denotes the normalized expansion rate in \ref{eq:fried}. The equation shows how the growth of perturbations through  gravitational instability is counter-acted by Hubble friction term due to the expansion of the Universe. For each chosen set of cosmological parameters determining the expansion rate \ref{eq:fried} we numerically solve eq. \ref{eq:linear_growth}. 

Inspection of eq. \ref{eq:fried} and \ref{eq:linear_growth} immediately shows the effect of introducing a negative $\Lambda$. 
At  early times $a\rightarrow 0$ the term related to matter density $\Omega_m\,a^{-3}$ dominates over all other terms, so that $H(a)$ (and hence $D(a)$) is almost independent on the other cosmological parameters. Analogously, in the late time regime $a\rightarrow 1$ the expansion rate in \ref{eq:linear_growth} - and hence the growth factor $D(a)$ - reduces exactly to the $\Lambda$CDM case since $\Omega_{x}+\Omega_{\Lambda}=1-\Omega_m\approx 0.7$, and again assuming a NCC has no effect on the growth of perturbations. 

However, at intermediate redshifts $z \sim 10-20$,  the value  of $\Omega_{\Lambda}$  appreciably affects $H(a)$; eq. \ref{eq:fried} shows that for decreasing values of $\Omega_{\Lambda}$ (and particularly for negative values)  a slower expansion rate $H(a)$ is obtained, resulting into larger growth factors in eq. \ref{eq:linear_growth}. Thus, at such redshifts, we expect a larger abundance of massive objects compared to $\Lambda$CDM, as shown in fig. 1 in \cite{menci2024excess}. 

Extending the above Press \& Schechter approach to derive the whole growth  history of the DM halos through subsequent merging of DM condensations (see, e.g., \cite{LC1993}, it is possible to derive the mass growth of the main branch in a DM growth history due to accretion and merging of haloes. This reads  (\cite{Neistein2006}, see, also \cite{LiuMassacc,Correa2015a}) 

\begin{eqnarray}
{1\over M(a)} {dM\over dt}=-{\sqrt{2\over \pi}}\,{1\over \sqrt{\sigma^2(M/q)-
\sigma^2(M)}}\,{d[\delta_c/D(t)]\over da}
\label{eq:mainprog_growth}
\end{eqnarray}
where $q$ can be computed to be in the range $q=[2.1-2.3]$ for flat cosmologies, with an uncertainty which is an intrinsic property of the extended Press \& Schecter theory. The comparison with up-to-date N-body simulations \citep{LiuMassacc} showed that the above predictions match the simulations results with typical values for the residuals $\lesssim 0.2$.
The authors in \cite{Correa2015a} showed that in the high-redshift regime we are interested in this paper, the above rate can be written as 
\begin{eqnarray}
{1\over M} {dM\over dt}=A\,f(M_0)\,E(a)
\label{eq:mainprog_growth_2}
\end{eqnarray}
leading to an exponential growth of the halo mass (as already obtained by \cite{wechsler2002}). When the mass $M$ is expressed in units $10^{12}\,M_{\odot}$ the normalization takes the value $A=10^2\,h\,yr^{-1}$, and $f(M_0)=1/\sqrt{\sigma^2(M_0/q)-\sigma^2(M_0)}$ depends on the final mass $M_0$ of the parent halo at $z=0$. We have verified that our mass growth histories (computed with small time steps according to the prescriptions in \cite{Neistein2006}) agree with those obtained in the above mentioned papers. 

\subsection{The growth of black holes and stars in the host dark matter halos}
\label{sec:ncc}

To compute the growth of the BHs and stellar mass hosted in the evolving DM halos we adopt a simplified approach: the aim is not to achieve detailed and precise predictions for the abundance of supermassive BHs, but rather to derive the  maximal BH growth that can be achieved in different NCC cosmologies under the most conservative assumptions for the physics of BH accretion and star formation. 

To this aim, we first consider a grid of halo masses $M$ at the initial redshift $z_i=20$, for which we compute the growth of the halos according to eq. \ref{eq:mainprog_growth}. This allows us to represent the  average mass growth of each mass bin.  Following an approach widely used in the literature (see, e.g., \cite{VolonteriHM2003,Barausse2012,Dayal2019}) we
 populate all halos collapsing from the large $\gtrsim 3.5\sigma $ peaks of the primordial density field \citep{MadauRees2001,VolonteriHM2003}; in the $\Lambda$CDM case, this corresponds to populate halos with mass $M\geq  1.1\,h^{-1}\,10^7\,M_{\odot}\,[(1+z)/20]^{-3/2}$ \citep{VolonteriHM2003}. 

Since we are interested in computing the maximal BH mass growth in the light-seeds scenario, we start from BH seeds with mass $M_{seed}=10^2\,M_{\odot}$.  From such an initial mass, BHs can grow both by accretion and merging. Since our aim is to derive the maximal SMBH growth that can be achieved in different NCC cosmologies, we assume continuous accretion at the Eddington limit 
\begin{equation}
M'_{ed}(a) \equiv \Bigg({dM_{BH}\over da}\Bigg)_{Edd}= {1\over \dot a}\,{4 \pi G\,M_{BH}(t) m_p\over \sigma_T \epsilon_r c}
\end{equation}
where $G$ is the gravitational constant, $m_p$ is the proton mass, $\sigma_T$ is the Thomson scattering optical depth, $\epsilon_r$ is the BH radiative efficiency and $c$ is the speed of light. 
The BH mass accreted in the timestep $\Delta a$ (in terms of the expansion factor $a$) as $M_{BH}(a+\Delta a) = M_{BH}(a)+(1-\epsilon_r) M'_{ed}(a) \cdot \Delta a$, where we assume a fiducial value $\epsilon_r=0.1$ for the radiative efficiency of accreting BHs. 
This allows us to compute the BH mass in each DM halo at the next time step. 

In the spirit of maximizing the BH growth in a given cosmological scenario, we assume an  instantaneous merging scenario. We assume that BH masses merge following the same relative rate of their host halos given by eq. \ref{eq:mainprog_growth}, i.e., that BH merging promptly follows that of their host halos. After updating the BH masses hosted in each DM halo after the above assumption, we iterate the procedure to get the average BH mass $M_{BH}$ hosted in halos with DM mass $M$ at a cosmic epoch $t$, up to a final time corresponding to $z_f=4$. This procedure is strictly valid for the main branch of the merging histories of DM halos. However, here we focus on the high redshift evolution of DM halos (and SMBHs hosted therein), when, due to the extremely rapid merging events, the mass growth of halos can be described on average by the continuous mass accretion onto a main branch in eq. \ref{eq:mainprog_growth}.  

As for the average star formation rate associated to DM halos, 
we adopt a phenomenological  representation of our knowledge about galaxy formation before the JWST era, thus assuming that the same physical processes  that shape the baryon conversion into stars at $z\lesssim 10$ are also driving  star formation at higher redshifts. 
Following the approach in \cite{menci2024excess}, the DM mass $M$ is  related to the star formation rate of galaxies  by $\dot m_* = \epsilon (M)\,f_b\,M$, where $f_b$ 
 is the cosmic baryon fraction. The efficiency $\epsilon(M)$ for the conversion of baryons into stars is taken from   \cite{Mason2015}  (see their fig. 1). This is a redshift-independent relation characterized by a maximal efficiency at masses $M\approx 10^{12}\,M_{\odot}$, and constitutes  a phenomenological  representation of our knowledge about galaxy formation before the JWST era. The star formation rate is related to the UV luminosity $L$   through the relation $\dot m_*/{\rm M_{\odot}\,yr^{-1}}=k_{UV}\,L/\,{\rm erg\,s^{-1}\,Hz^{-1}}$ with $k_{UV}=0.7\,10^{-28}$  assuming a Chabrier initial mass function \citep{Madau2014ApJ},  so that $L\propto \epsilon(M)\,M$. From the star formation rate above, the stellar mass $m_*$ associated to a DM halo $M$ can be computed analytically. Instead of performing  time integration of $\dot m_*$ over the progenitors, we adopt a simple phenomenological approach proposed by \cite{BehrooziSilk2015}  based on observed average relations. Measurements of the $M_*$-$M$ relation at $z=4-10$ show that their relation can be approximated as a power law  $M_*=M_{*0}\,(M/M_0)^{\alpha}$ for DM masses 
$M\leq 5\times10^{11}\,M_{\odot}$ considered here, see \cite{BehrooziSilk2015,Behroozi2019,Behroozi2020}. 
As noted by \cite{BehrooziSilk2015}, this implies that the stellar and DM specific accretion rates are related by $\dot M_*/M=\alpha\,\dot M/M$. The stellar mass $M_*$ can then be computed from $\dot M_*$ and from the DM specific accretion rate in eq. \ref{eq:mainprog_growth} from the observed values of $\alpha$, which are within the range $\alpha=[1-1.5]$ \citep{BehrooziSilk2015,Behroozi2019,Behroozi2020}). The uncertainties in $M_*$ associated to the above range of values for $\alpha$ is considered when we derive the stellar mass $M_*$. 

In the following, all distributions $dN(X,z)/dX=\Big[dN(M,z)/dM\Big]\,(dM/dX)$ of any of the above observables $X$ (i.e., the UV luminosity of galaxies, their star formation rate, stellar mass, BH mass and AGN luminosity) will be derived from the 
DM halo mass function \ref{eq:dndm} after connecting the 
halo mass $M$ to the observable $X$ following the modeling  
described above. 

\subsection{The cosmological framework}
\label{sec:cosm_frame}

In our previous papers \citep{Menci_2024,menci2024excess}
 we have shown that quintessence models  (with $w_0\geq -1$ and $w_a\geq 0$) with negative $\Omega_{\Lambda}$
  can provide a solution to the excess of both massive galaxies at $z\gtrsim 6$ and of luminous galaxies at $z\gtrsim 10$ observed by JWST. This is due to the 
accelerated growth of cosmic structures, as seen in fig. 1 in \cite{menci2024excess}. In addition, such models do not yield crossing of the  phantom regime, so that $w$ remains in the non-phantom regime $w\geq -1$ throughout the cosmic history.

Since our aim here is to show that such cosmologies can provide a  simultaneous solution to both the above excess  and to the early growth of supermassive BHs at high redshifts (without invoking heavy seeds or Super-Eddington accretion), in the following we shall focus on such combinations. We further restrict the parameter space by requiring their consistency 
with the existing cosmological constraints. To this aim, we consider the following datasets:
\newline
i) latest data release from the Atacama Cosmology Telescope ACT  DR6 \citep{AtacamaCosmologyTelescope:2025blo} with a ${\it Planck_{cut}}$ likelihood from Planck PR3 data release \cite{Planck:2018vyg} (for details of the data combination, see \cite{mukherjee2025quintessentialimplicationspresenceads} and references therein).
\newline
ii) the DESI-BAO DR2 \citep{DESI:2025zgx} measurements for the redshift range $0.1 < z < 4.2$.
\newline
iii) the DES-SN5YR \citep{Abbott_2024, DES:2024upw} sample for SNIa luminosity measurements in the redshift range $ 0.1 < z < 1.13$ complemented by low-$z$ SNIa measurements in the redshift range $0.025 < z < 0.1$.

We do not consider constraints from reionization. Although recent works in the literature \citep{chakraborty2025} have shown that the measurements of the cosmic reionization history may significantly s
restrict the parameter space of DDE with negative cosmological constant, strongly limiting their boost to structure formation, 
 the modeling of reionization is still affected by appreciable uncertainties. For instance, in the above paper, only the contribution of galaxy UV emission is considered as a source of reionization while, especially in view of our results  below, a significant fraction of UV ionizing flux can result from AGN emission. 

We run Markov Chain Monte Carlo (MCMC) methods to derive constraints on our set of cosmological parameters (for a full discussion of the methodology and the resulting constraints, we refer the reader to~\cite{mukherjee2025quintessentialimplicationspresenceads}.

\begin{figure}[!ht]
\includegraphics[width=0.8\linewidth]{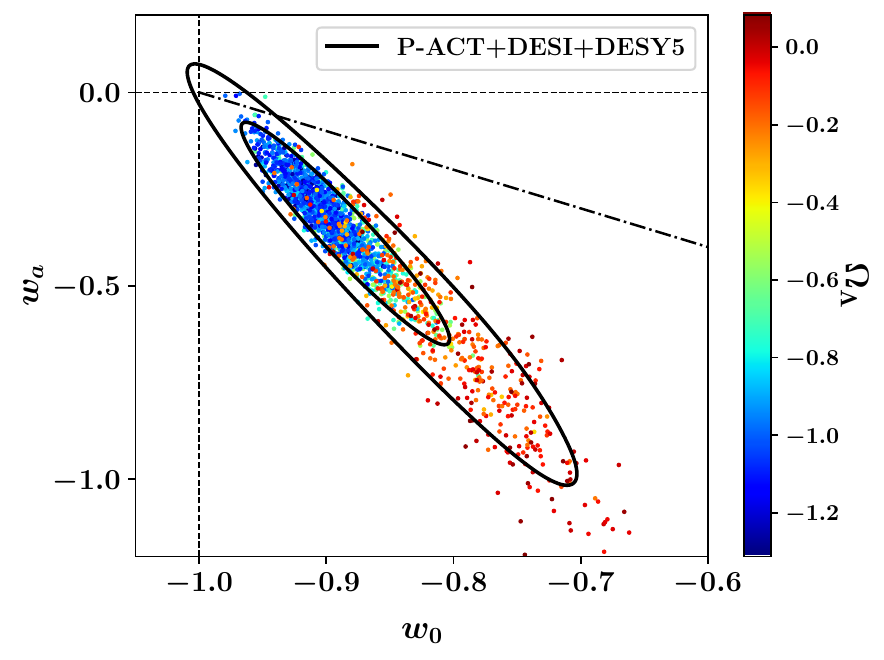}
\caption{2D confidence contour in the $w_{0}-w_{a}$ plane for our model using CMB+DESI+DES data combination. The Dash-Dotted line is phantom line. Above this line, the allowed region is always non-phantom, where as below this line, the allowed region is early phantom and late non-phantom. Dots represents different Monte Carlo realizations. The color code shows the value of $\Omega_{\Lambda}$ that allows for consistency with the above cosmological probes}
\label{w0wa_contours}
\end{figure}

In Fig. ~\ref{w0wa_contours}, we show the allowed region in the $w_{0}-w_{a}$ plane for the data combination mentioned above. As one can see, there is small but finite region in the parameter space which is always non-phantom, and where $\Omega_{\Lambda}$  is negative. 

\begin{figure}[!ht]
\includegraphics[width=1.\linewidth]{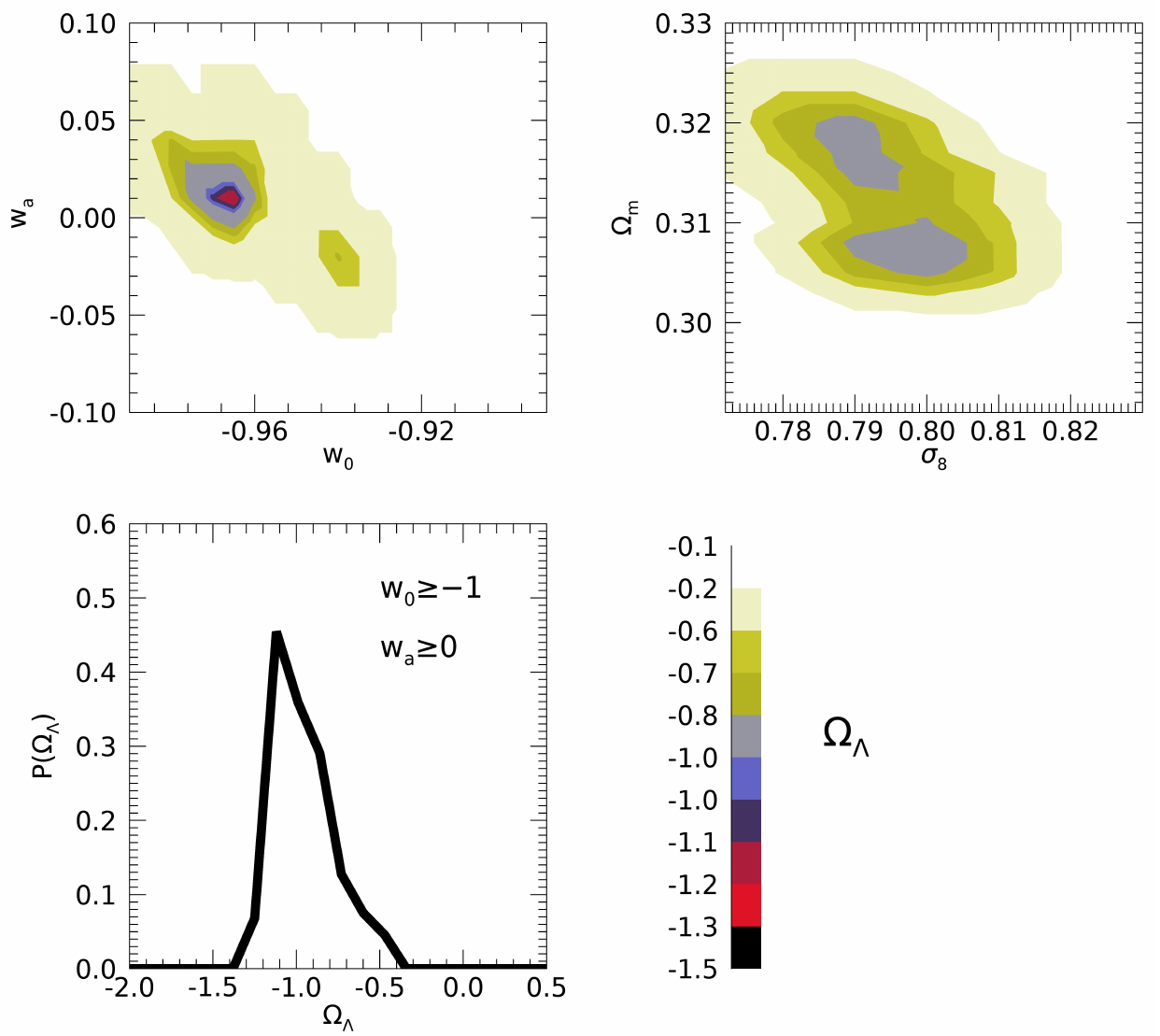}
\caption{Top-Left Panel: Based on our Monte Carlo procedure for the selection of  ($w_0$-$w_a$, $\Omega_{\Lambda}$) combinations combinations consistent (at 2-$\sigma$ level) with the considered  cosmological probes, we show as colored contours the average $\Omega_{\Lambda}$ as a function of ($w_0$-$w_a$). 
The color code is shown by the bar in the bottom-right of the figure. \newline
Top-Right Panel: for the combinations ($w_0\geq -1$, $w_a\geq 0$) considered here, we show the values of $\Omega_m$ and $\sigma_8$ consistent (at 2-$\sigma$ level) with the considered cosmological probes. The 
color code refers to the average values of $\Omega_{\Lambda}$ corresponding to each combination, as shown by the bottom bar.\newline
Bottom-Left Panel: the probability distribution of $\Omega_{\Lambda}$ resulting from our Monte Carlo procedure for the combinations ($w_0\geq -1$, $w_a\geq 0$) considered here.
}
\label{cosmo_contours}
\end{figure}

We explore the properties of such a region in a closer detail in Fig.  ~\ref{cosmo_contours}, where we focus on non-phantom combinations ($w_{0},w_{a}$) that match  (within $2\sigma$) the cosmological observations above. In the top-left panel, 
we show the average values of $\Omega_{\Lambda}$ corresponding to such ($w_{0},w_{a}$) combinations. In the  
upper-right panel of Fig. ~\ref{cosmo_contours} we show the values of $\sigma_8$ and of $\Omega_m$, with the average value of $\Omega_{\Lambda}$ associated the non-phantom region of the $w_{0},w_{a}$ plane. For the same region, the whole distribution of $\Omega_{\Lambda}$ values resulting from our Monte Carlo procedure is shown in the bottom-left.

\begin{figure}[!ht]
\includegraphics[width=0.9\linewidth]{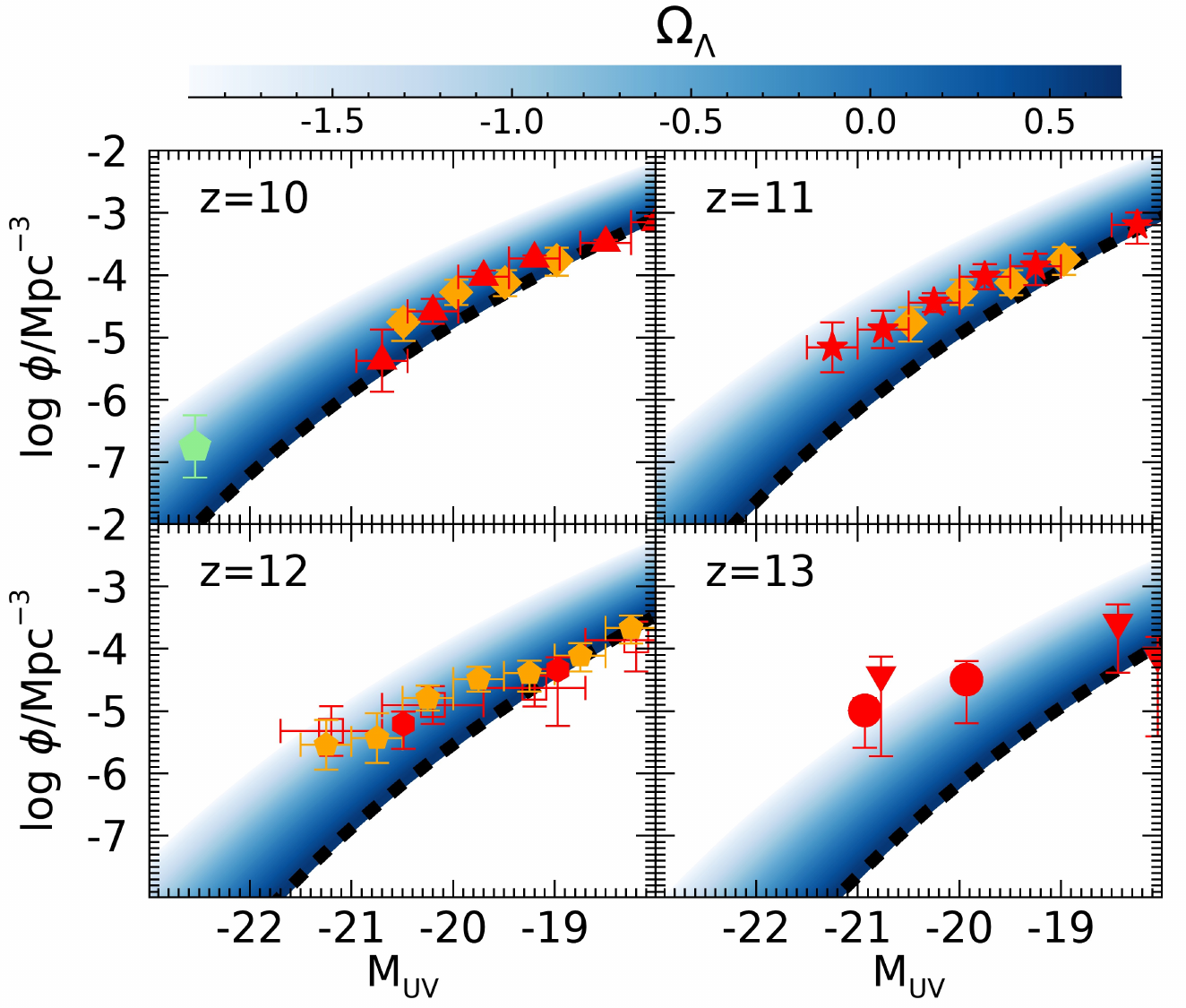}
\caption{The high redshift UV luminosity functions of galaxies for our fiducial combination of ($w_0$, $w_a$) in the different redshift bins shown in the labels. 
The color code corresponds  the different values of $\Omega_{\Lambda}$ as shown in the top bar, while the dashed line mark the results for the standard $\Lambda$CDM cosmology. 
We compare with measurements from \cite{Finkelstein_2023} (diamonds) \cite{Donnan_2023} (stars) \cite{McLeod_2023a} (upward triangle)
\cite{Harikane_2023} (open square) \cite{adams24} (pentagon) \cite{Bouwens_2022} (circle) \cite{Robertson_2024} 
 (downward triangle). 
}
\label{LF_JWST}
\end{figure}

Based on the results above, we shall consider here a fiducial choice $w_0=-0.98$ $w_a=0.08$ for the equation-of-state parameter, allowing $\Omega_{\Lambda}$ to assume all values in the distribution shown in the bottom-left panel of fig. ~\ref{cosmo_contours}. This ensures that the model considered here can be made consistent with the observed abundance of massive galaxies at $z\gtrsim 6$ (see \cite{Menci_2024}) and with the observed luminosity function of bright  galaxies measured by JWST at $z\gtrsim 10$ that we show  in Fig.  ~\ref{LF_JWST}. We have checked that our results do not differ appreciably when other combinations ($w_0$,$w_a$) are considered among the values defining  the  non-phantom regime ($w_0\geq -1$ - $w_a\geq 0$) for the equation-of-state parameter represented in the 
upper right quadrant of Fig. ~\ref{cosmo_contours}. 
Inspection of Fig. \ref{LF_JWST} shows that values of $\Omega_{\Lambda}\approx -1$ can yield an abundance of high-redshift bright galaxies close to the observed values.

\section{Results}

\begin{figure}[!ht]
\includegraphics[width=0.95\linewidth]{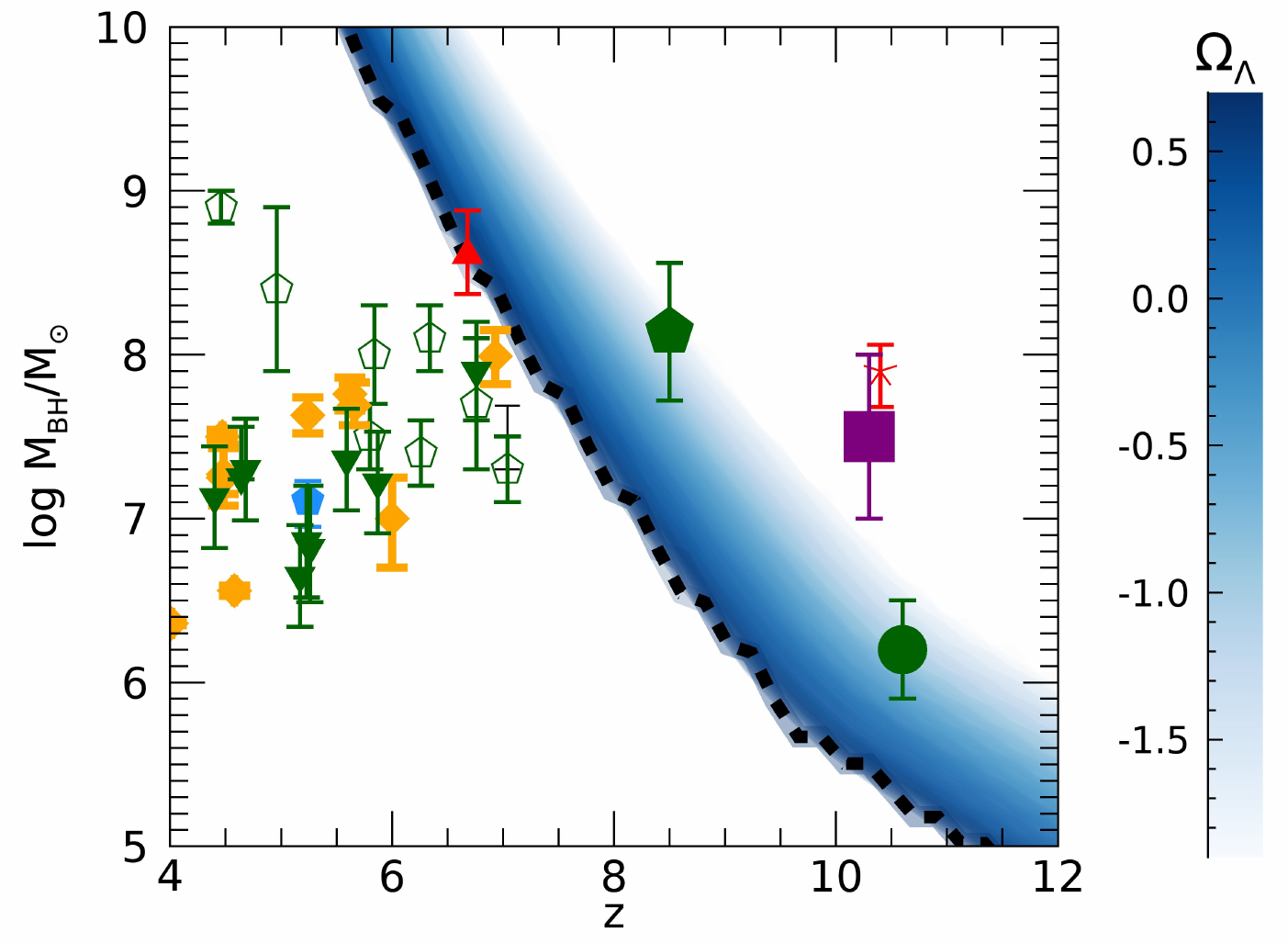}
\caption{The maximal growth of BHs from  $M_{seed}=10^2\,M_{\odot}$ at $z_{seed }= 25$ is shown for the different values of $\Omega_{\Lambda}$ shown in the color bar for our fiducial combination $w_0=-0.98$, $w_a=0.08$; the dashed line mark the result for the standard $\Lambda$CDM cosmology.  To maximize the BH mass we assume continue Eddington-limited accretion. We compare these models to  black holes measured by \cite{Bogdan_2024NatAs} (solid square) \cite{Furtak2024Nature} (cross)  \cite{Greene2024ApJ} (empty petagons) \cite{Juod_balis_2024} (upward triangle) \cite{Harikane_2023} (filled diamonds) \cite{Maiolino2024a} 
(downward triangles) \cite{Maiolino2024Nature} (solid circle)  \cite{Kocevski2023ApJ} (solid pentagon)
}
\label{BHgrowth}
\end{figure}

Based on the above  framework, we proceed to compute the  growth of supermassive BHs in the framework of NCC cosmologies. 
As discussed in Sect. 3.2, our aim is not to achieve detailed and precise predictions for the abundance of BHs, but rather to derive the  maximal BH growth that can be achieved in different NCC cosmologies under the most conservative assumptions for the physics of BH accretion and star formation. Following the above discussion, we populate all halos collapsing from the large $\gtrsim 3.5\sigma$ peaks of the primordial density
 with  initial BH seeds with mass $M_{seed}\approx 10^2\,M_{\odot}$, and follow their evolution as described in Sect. 3.2 assuming continuous Eddington-limited accretion. Under the assumptions above, the  maximal growth of BHs hosted in
halos corresponding to $3.5$-peaks of the density field is shown in Fig. \ref{BHgrowth} for our fiducial cosmological framework for different values of  $\Omega_{\Lambda}$.

As obtained in previous works in the literature (see, e.g., \cite{Dayal2024A&A} and references therein), in the $\Lambda$CDM cosmology with light BH seeds $M_{\odot}=10^2\,M_{\odot}$, only BHs with mass $M_{BH}\lesssim 10^5\,M_{\odot}$ can be 
in place by $z\approx 10$, and  $M_{BH}\lesssim 10^8\,M_{\odot}$   by $z\approx 8$, even under the assumption of continuous Eddington accretion. 

On the other hand, assuming NCC cosmologies yields an accelerated growth of BHs at $z\geq 10$ (see \ref{eq:linear_growth}) so as to allow for the presence of  BHs as massive as $M_{BH}\sim 10^7\,M_{\odot}$ at $z\sim 10$
 for $\Omega_{\Lambda}\approx -1$ without the need for Super-Eddington accretion or for heavy BH seeds. This shows that NCC cosmology can constitute an alternative solution to the problem posed by the presence of massive BHs at early cosmic epochs.

Motivated by such results, we proceeded to investigate whether the same cosmology can account for the other problem posed by the recent JWST observations, namely, 
the large number density of AGNs identified by JWST, both as type- I/type-II AGNs (see e.g. \cite{HaricaneAGN2023ApJ,Maiolino2024a,Scholtz2025A&A}) and as LRDs (see e.g. \cite{Matthee2024,Greene2024ApJ,Kokorev_2024,Akins2024}). These studies show that the number density of JWST-detected systems lies between 1 and 2 dex larger than the extrapolation of the UV quasar luminosity function at 
$z\gtrsim 6$ \citep{shen2020} and higher than the estimated density from deep X-ray observations \citep{giallongo2015,giallongo2019}. Such large number densities exceed by $\sim 1$ order of magnitude the predictions of $\Lambda$CDM models implementing Eddington-limited models for the BH accretion (see \cite{TrincaAGN} and references therein). 

\begin{figure}[!ht]
\includegraphics[width=1\linewidth]{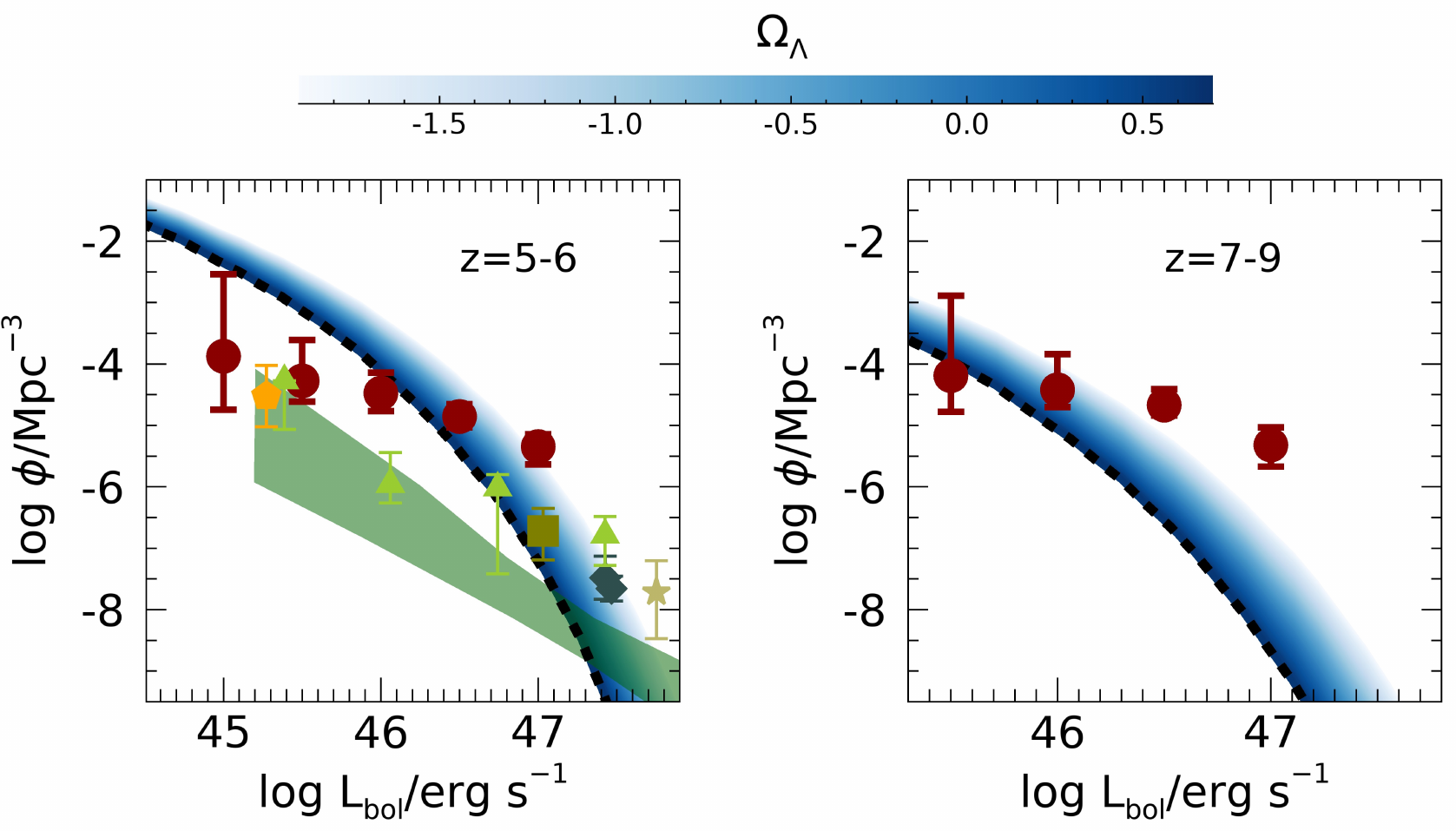}
\caption{The bolometric luminosity function of AGN under the assumption of continuous accretion at the Eddington rate   is shown for the different values of $\Omega_{\Lambda}$ shown in the color bar for our fiducial combination $w_0=-0.98$, $w_a=0.08$; the dashed line mark the result for the standard $\Lambda$CDM cosmology.  We compare with the measurements by \cite{Akins2024} based on JWST observations (red circles), derived under the assumption that the  emission of LRD is entirely contributed by AGN, and by \cite{Harikane_2023} (orange pentagon). For comparison, we also show the pre-JWST measurements by  \cite{shen2020} as a shaded area, by \cite{Glikman} (square), \cite{Grazian2023} (diamonds), and the X-ray measurement through Chandra COSMOS data by \cite{Barlow} (triangles) and \cite{barlow2023} (star). 
}
\label{Lbol_agn}
\end{figure}

Figure \ref{Lbol_agn} shows the AGN bolometric luminosity function computed in different NCC models at $z=5-7$ and $z=7-9$. 
Again, our aim is not to test a detailed description of AGN accretion, but rather to show the  maximal abundance of AGN that can be achieved in any given cosmological model under the assumption of Eddington-limited accretion and BH seeds with mass $10^2\,M_{\odot}$. The model results are compared with the recent estimates by \cite{Akins2024} based on JWST observations; the pre-JWST measurements  by  \cite{shen2020}  are also shown for comparison. 

As obtained in previous works in the literature (see, e.g. \cite{TrincaAGN}), 
in the Eddington-limited accretion scenario the predicted bolometric luminosity functions predicted in the $\Lambda$CDM cosmology drops by more than two orders of magnitude below the number density estimated by  \cite{Akins2024} at the bright-end of the distribution. 
 Also in this case, NCC models alleviated the discrepancy between observations and predictions, although the brightest point at $z\geq 7$ is still under-predicted. 

However, the observational bolometric LF \citep{Akins2024}  reported here  relies on the assumption that LRD emission is  dominated by AGN. In fact, several studies suggest that both the stellar and the AGN components contribute to the SED (see e.g. \cite{volonteri2024,taylor2024}) thus resulting in a lower estimated AGN luminosity. Although this could bring the observed AGN luminosity function in better agreement with models (including those within the $\Lambda$CDM framework), this would result in a  further additional component  contributing to the stellar mass function and hence appreciably worsen the problems 
 of current galaxy formation models in matching the abundance of massive galaxies at high redshifts $z\gtrsim 6$. 

\begin{figure}[!ht]
\includegraphics[width=1\linewidth]{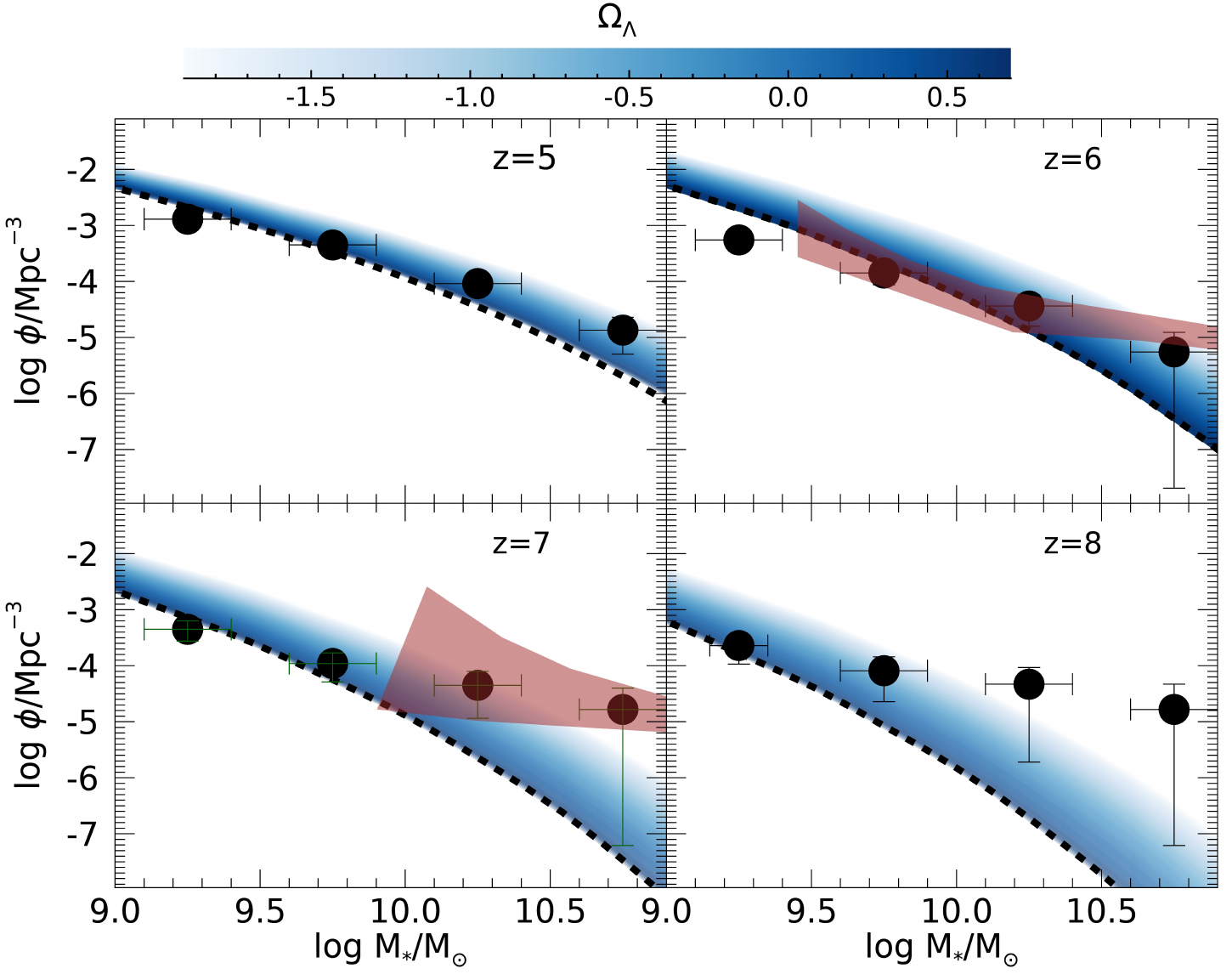}
\caption{The stellar mass function of galaxies for the different values of $\Omega_{\Lambda}$ shown in the color bar for our fiducial combination $w_0=-0.98$, $w_a=0.08$; the dashed line mark the result for the standard $\Lambda$CDM cosmology.  We compare with the JWST measurements by \cite{Weibel2024} with the estimate by \cite{Akins2024} (shaded area) who considered the  additional contribution by LRD  under the assumption that their emission is entirely contributed by stars. 
}
\label{nmstar}
\end{figure}

This is illustrated in fig. \ref{nmstar} where we compare the stellar mass function of galaxies in different cosmologies with the JWST measurements by \cite{Weibel2024}. We show the increase in the abundance of massive galaxies that would be obtained if the emission of all LRD  in the sample by \cite{Akins2024} was contributed by stars. In this extreme case a strong excess relative to the $\Lambda$CDM predictions would exist at $z\approx 7$, while a much better agreement would be obtained for NCC models. Although such an excess is still to be confirmed by extending the portion of the sample which include MIRI measurements, this indicates that the tension with the $\Lambda$CDM predictions yield by the abundance of LRD would show up either in the luminosity function of AGN (at least under the assumption of Eddington-limited accretion) or in the stellar mass function. Future measurements, able to increase the accuracy of stellar mass measurement, will help to assess the issue. 

\section{Discussion and conclusions}
Motivated by recent observational indications for a dynamical behavior of the dark energy, we have explored the possible impact of assuming dynamical DE models with negative cosmological constant on the growth of massive BHs at high redshifts and on the abundance of AGN. We have shown that such cosmological models can account for the observed presence of  massive Black Holes with masses $M_{BH}\gtrsim 10^7\,M_{\odot}$ observed at $z\gtrsim 8$ and for the large over-abundance of AGN at $z\gtrsim 6$ with respect to pre-JWST, without the need to invoke super-Eddington accretion or massive BH seeds. 
As shown in previous results, the same negative cosmological constant (NCC) models can also account for the large
abundance of bright galaxies observed by JWST at $z\gtrsim 10$ and the abundance of massive galaxies at $z\gtrsim 6$, both largely exceeding the pre-JWST
expectation in a $\Lambda$CDM Universe.

We stress that each of the 
 above observations can in principle be explained by a variety of different astrophysical processes that should become dominant at high redshifts, as discussed in the Introduction. 
 Nevertheless, recent results from different cosmological probes suggest that a departure from $\Lambda$CDM toward Dark Energy models may be necessary.  Indeed, the indications for an emerging crack in the cosmological constant paradigm are increasing as new analyses of the DESI results (in combination with other cosmological probes) are performed with a variety of different techniques: e.g. 
 in \cite{chaudhary2025} 
 several DE scenarios are explored, along with cosmologies allowing for spatial curvature, using Metropolis-Hastings Markov Chain Monte Carlo techniques, finding strong statistical evidence for  dynamical DE  behavior; such a conclusion has been confirmed by analyses extended to consider a wider range of cosmological datasets including DESI DR2 BAO and Ly$\alpha$ data, CMB compressed likelihoods, Big Bang Nucleosynthesis, cosmic chronometers, and multiple Type Ia supernova by 
 \cite{capozziello2026}, and by analyses 
 adopting model-independent techniques to  compare the DESI DR1 and DR2 predictions for the $\Lambda$CDM model and dynamical DE model for different redshift ranges \citep{ChaudharyApJ2025}. The mounting indications of a departure from $\Lambda$CDM calls for the investigations of the impact of the emerging new cosmological scenarios on the interpretations of the JWST observations independently of the possible astrophysical  interpretations. Indeed, as showed here, such a new cosmological framework could already by itself account for the enhanced abundances of galaxies and AGN detected at high redshifts observed by JWST. Even if future observations would confirm a downward rescaling of the currently high AGN abundances or BH masses, such as in the ``Black Hole Envelope'' interpretation \cite[e.g.,][and references therein]{Umeda25}, NCC cosmologies would still represent a viable background framework to facilitate structure formation without invoking extreme values of the star formation efficiencies or too large initial BH seeds and/or too prolonged super-Eddington accretion phases.
More specifically, we found that 
 NCC models with non-phantom behavior, consistent with the results of cosmological probes can account for the JWST observations concerning the properties of the BH, AGN, and galaxy populations at high redshifts. We stress that the region of parameter space that we focused on (see Sect. 3.3) has been selected on the basis of consistency with cosmological probes only, i.e., without considering possible constraints from the observed cosmic reionization history. 
 Although a recent work  \citep{chakraborty2025} has pointed out that the region of the parameter space we have investigated can be significantly restricted when such constraints are included, such result was obtained under the assumption that reionization is entirely driven  by galactic emission; the assumed escape fraction adopted in such work is based on the same assumption. Here, however, we have shown that in NCC scenarios a relevant population of AGN is present at high redshift, with a duty cycle much larger than that of the local Universe, potentially contributing to the reionization, and thus relaxing the constraints on the parameter space. We plan to explore in detail the impact of the assumed cosmological scenario on the cosmic reionization history in a subsequent paper, taking into account the contribution of both galaxies and AGN, with the inclusion of a realistic model to relate the escape fraction in AGN with their feedback on the intergalactic gas. 

From a theoretical point of view, our results show a NCC Dark Energy models with a non-phantom equation of state constitute an attractive possibility: in fact, such models:   \\
\noindent
(i) are naturally motivated in the context of string theory\\
(ii) allow for consistency with latest CMB+BAO+Sn observations without invoking non-phantom behaviour for the dark energy \\
(iii) yield finite life time for the Universe with recollapse happening in the future as shown in \cite{mukherjee2025quintessentialimplicationspresenceads}.

Given that such a theoretically motivated and observationally consistent model can indeed explain the growth of massive BHs and abundances of AGN at high redshift as confirmed by recent JWST measurements, makes such a model highly attractive to pursue as a viable cosmological set up.

\label{sec:ncc}

\begin{acknowledgements}
INAF Theory Grant ''AGN-driven outflows in cosmological models of galaxy formation'', INAF Mini-grant ``Reionization and Fundamental Cosmology with High-Redshift Galaxies", and INAF GO Grant "Revealing the nature of bright galaxies at cosmic dawn with deep JWST spectroscopy". PM acknowledges funding from the Anusandhan National Research Foundation (ANRF), Govt of India, under the National Post-Doctoral Fellowship (File no. PDF/2023/001986). AAS acknowledges the funding from ANRF, Govt of India, under the research grant no. CRG/2023/003984. PM and AAS acknowledge the use of the HPC facility, Pegasus, at IUCAA, Pune, India.
\end{acknowledgements}

\bibliographystyle{aa}

\bibliography{biblio}

\end{document}